# Artemis-enabled Stellar Imager (AeSI):
# A Lunar Long-Baseline UV/Optical Imaging Interferometer

Gioia Rau[1,2], Kenneth G. Carpenter[2], Tabitha Boyajian[3], Michelle Creech-Eakman[4], Julianne Foster[5], Margarita Karovska[6], David Leisawitz[2], Jon A. Morse[7], David Mozurkewich[8], Sarah Peacock[9,2], Noah Petro[2], Paul Scowen[2], Breann Sitarski[2], Gerard Van Belle[10], Erik Wilkinson[5]

[1]NSF, Alexandria, VA, USA; [2]NASA Goddard Space Flight Center, Greenbelt, MD, USA; [3]LSU, Baton Rouge, LA, USA; [4]NMT/MROI, Socorro, NM, USA; [5]BAE Systems, Boulder, CO, USA; [6]CfA | Harvard & Smithsonian, Cambridge, MA, USA; [7]Caltech, Pasadena, CA, USA; [8]Seabrook Engineering, Lanham, MD, USA; [9]UMBC, Baltimore, MD, USA; [10]Lowell Obs., Flagstaff, AZ, USA

**ABSTRACT**

NASA's return to the Moon presents unparalleled opportunities to advance high-impact scientific capabilities. At the cutting edge of these possibilities are extremely high-resolution interferometric observations at visible and ultraviolet wavelengths. Such technology can resolve the surfaces of stars, explore the inner accretion disks of nascent stars and black holes, and eventually enable us to observe surface features and weather patterns on nearby exoplanets. We have been awarded Phase 1 support from NASA's Innovative Advanced Concepts (NIAC) program to explore the feasibility of constructing a high-resolution, long-baseline UV/optical imaging interferometer on the lunar surface, in conjunction with the Artemis Program. A 1996 study comparing interferometers on the Moon versus free-flyers in space concluded that, without pre-existing lunar infrastructure, free-flyers were preferable. However, with the advent of the Artemis Program, it is now crucial to revisit the potential of building lunar interferometers. Our objective is to conduct a study with the same level of rigor applied to large baseline, free-flying interferometers during the 2003-2005 NASA Vision Missions Studies. This preparation is essential for timely and effective utilization of the forthcoming lunar infrastructure. In this paper, we highlight the groundbreaking potential of a lunar surface-based interferometer. This concept study will be a huge step forward to larger arrays on both the moon and free-flying in space, over a wide variety of wavelengths and science topics. Our Phase 1 study began in April 2024, and here we present a concise overview of our vision and the progress made so far.

## 1. INTRODUCTION

We are developing a groundbreaking concept, funded by NASA's Innovative Advanced Concepts (NIAC) program, to deploy a UV/optical imaging interferometer on the lunar surface. The Artemis-enabled Stellar Imager (AeSI) would provide unprecedented high-resolution images of astronomical objects, including, but not limited to: stars, accretion disks, Active Galactic Nuclei (AGN), stellar magnetic activity, and the host stars in exoplanetary systems. A study from 1996 [1] favored space-based interferometers due to the lack of lunar infrastructure; but with the foreseeable development of such infrastructure through the Artemis Program [2], there's a renewed interest in exploring interferometers on the lunar surface, leveraging previous studies like the 2003-2005 Vision Mission Study for the free-flying Stellar Imager (SI) [3].
The construction of interferometers beyond the surface of the Earth is technically challenging and will benefit enormously from smaller precursor or full-up missions on the lunar surface, where a mix of human and, potentially, robotic support will be available to build, debug, and maintain it. A schematic overview of our mission concept study is shown in Figure 1.

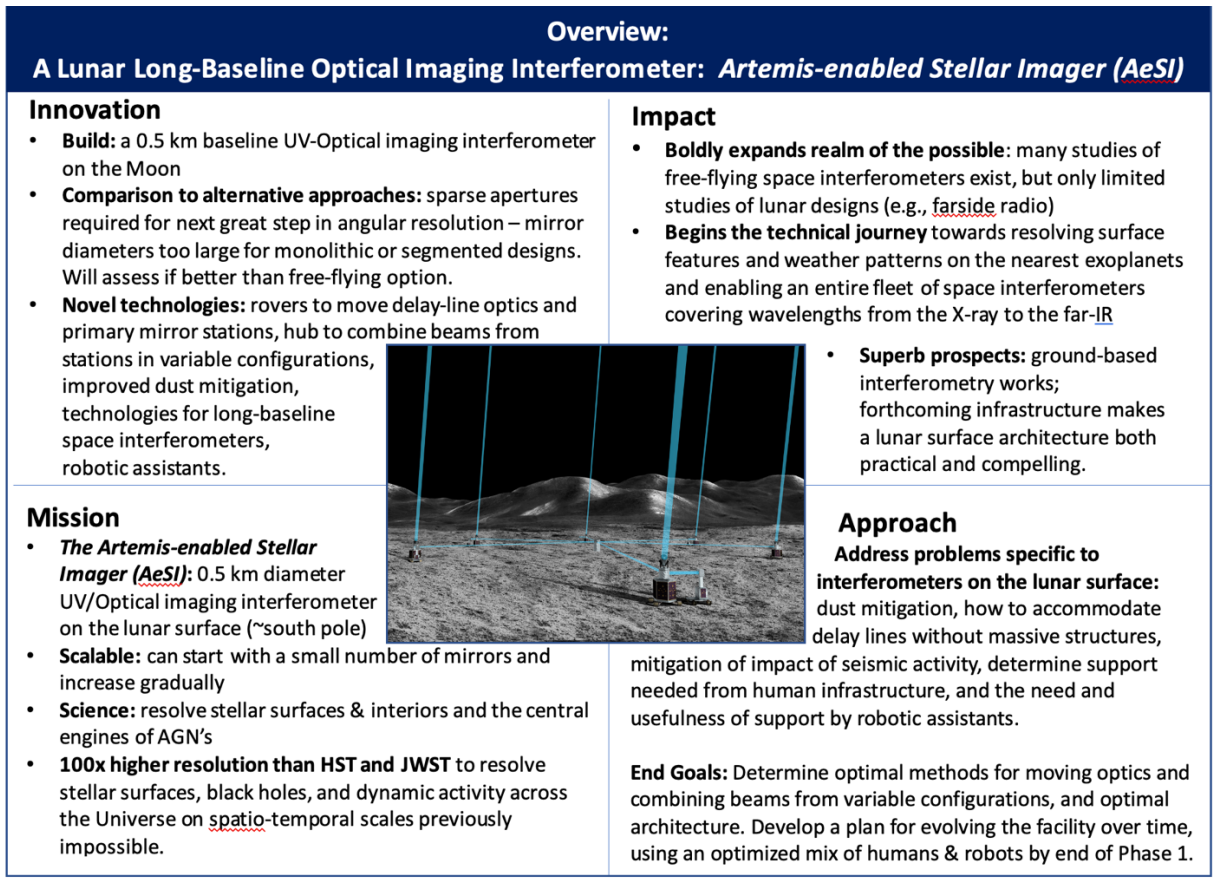

Figure 1. A summary chart showing the innovations and approach for enabling our mission concept, as well as the expected impact of such a mission.

## 2. SCIENCE CASE

AeSI will offer unparalleled capabilities in high-resolution imaging of celestial bodies. The groundbreaking science enabled includes, but is not limited to: 1) imaging the surfaces of nearby (<4pc) solar-type stars and more distant supergiants (>2kpcs), to observe magnetically driven activity such as plages, starspots, and convection. Such studies will improve our understanding of the solar/stellar dynamo and our ability to forecast future solar/stellar activity cycles and their impact on Earth and on the habitability of extrasolar planets; 2) imaging accretion disks around nascent stars; and 3) imaging the central engines of Active Galactic Nuclei (AGN).

AeSI's ability to enable time-resolved views of stellar surfaces and, via spatially resolved asteroseismology, to probe the interior of stars, will revolutionize our understanding of solar and stellar magnetic activity and of solar/stellar dynamos, offering crucial insights into space weather and the habitability of exoplanets (see Figure 2).

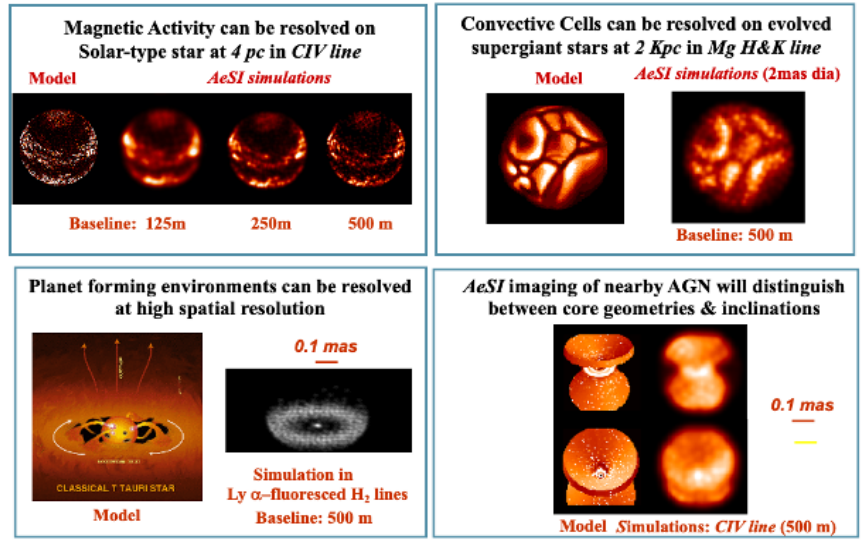

Figure 2. Simulations of AeSI observations showing the improvement in imaging the surfaces of sun-like stars with increasing baselines (array diameters), and the expected images of supergiants, planet-forming environments, and AGN cores with 500m baselines.

## 3. NIAC STUDY SCOPE, CHALLENGES, AND TECHNICAL APPROACH

We are studying the feasibility of building a 0.5 km diameter long-baseline UV-Optical interferometer (AeSI) on the lunar surface, utilizing supporting infrastructure to be provided by the Artemis human exploration program. AeSI can extend what can be done with Earth-based arrays thanks to its ability to perform in the UV. This study will provide an assessment of the relative costs, performance, ease of construction, deployment, operations and maintenance of free-flyers vs. moon-based designs and provide a path forward for the future of space interferometry.

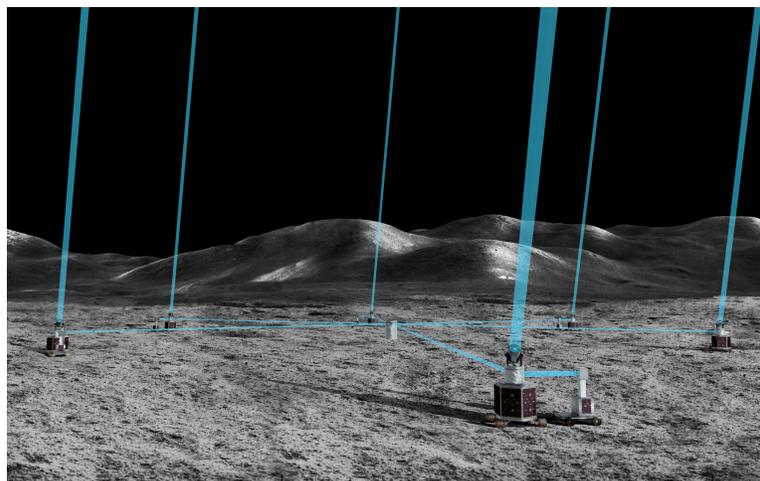

Figure 3. Artist's Concept of one possible initial six-element design, utilizing rover-based primary stations along with independent delay-line stations for each element. Other options include placing the primary stations on rails (e.g., like the Very Large Array - VLA) and/or including delay-lines within each station and taking out long-delays by altering the array configuration. The initial concept could be later expanded to 30, or more, elements.

AeSI requires solving challenges such as: designing long and variable delay lines on the lunar surface, accommodating beams from mirror arrays, mitigating dust and molecular contamination [4, 5, 6] and seismic activity [7], and identifying the support needed from lunar infrastructure. We propose to adapt techniques demonstrated on Earth-based interferometers while addressing these lunar-specific challenges.

There are many significant technical challenges to putting an interferometer on the Moon, but all of them have solutions that could be implemented as soon as the Artemis infrastructure is available. We describe here a few of these, along with potential solutions. More in-depth descriptions of challenges and potential solutions will be given in the Phase 1 NIAC final report. Optical interferometers have been operational on Earth since the 1990's [8,9,10,11]. Building a telescope on the Moon requires optics, actuators, sensors, a control system, and power, all of which have space heritage. All these systems can easily be deposited on the Moon with a lunar lander, such as the SpaceX Starship Human Landing System (HLS), which offers ample capacity in terms of both downmass and volume, with its cargo downmass of ~100 metric tons, and a fairing of 9m x 18m. This will allow us to deliver in the future far more than the 13 pieces (6 primary mirrors + 6 delay line rovers +1 beam combiner) of an initial 6-element interferometer, allowing us to expand the interferometer to potentially a very large one, which could include 30+ elements. The lunar surface is harsher due to larger temperature swings and the presence of dust. Heaters and sun shields can control temperature swings. Dust, and perhaps molecular contamination for the UV observations, may be another challenge; however, accommodations can be made. Retro-reflectors have been working on the lunar surface for 50 years losing only 5-6% reflectivity per year [12] and beginning in 2022 a whole new era of sophisticated instruments has been deployed on the lunar surface and the scope of the dust problem and how to deal with it will soon be much better known (e.g., [13]). We are studying how to prevent the dust from accumulating: whether to use automatic systems or use on-site personnel/robotics to clean the optics once or twice per year. We will learn further details regarding specific surface vibration environment in the south polar region from: (i) the *JPL FSS seismometer* [14] for the 2025 Schrodinger Basin landing; and (ii) the Chandrayaan-3 mission that landed near the South Pole, which will enable us to optimize our seismic activity mitigation procedures and designs.

Given this background, our primary approach to technical challenges involves adapting proven Earth-based techniques, while simultaneously innovating to develop space-durable solutions. Our team will focus on the following innovations, to create a working interferometer on the lunar surface:

1) Assess the **science case** for a UV/optical interferometer in space. This has been made very compellingly in the SI Vision Mission (VM) Report [15] and it's becoming increasingly evident that the moon offers unparalleled scientific opportunities [16]. However, we must refine the scientific objectives and our Design Reference Mission (DRM) to optimize the overall program to accommodate the unique constraints of lunar surface observation. These constraints encompass a spectrum from lunar geography to power challenges, and more limited sky accessibility.

2) Flow down the science requirements into **mission design parameters**.

3) Perform **engineering and architecture studies** implied by the required innovations and challenges.

4) Identify the support needed from **lunar infrastructure**.

5) Develop a **plan** for maintaining and evolving the facility over time, using an optimized mix of humans and robots. Compare the costs and difficulties for building a full-up array with up to 30 primary elements all at once vs. starting with a smaller array of ~6 elements and adding elements over time. For the latter, consider whether the original beam-combining hub, containing the detector and the observatory, should be compatible with the full array, or be upgraded with time.

## 4. SUMMARY AND IMPACTS

The urge to expand our understanding of the Universe, both near and far, has always driven the desire for higher sensitivity and higher angular resolution on the sky. Monolithic and segmented telescopes can only take us so far in the quest for higher angular resolution. The most compelling science cases now push for resolution far beyond what those telescope designs can provide, requiring apertures 100's of meters to kilometers in diameter. These require dispersed aperture, interferometric designs, making interferometers like *AeSI* the future of astronomy (see [15, 16]).
The science enabled by high-angular resolution interferometry is compelling at all wavelengths, including: the study of black hole event horizons in the X-ray, the study of stellar magnetic activity, star-planet interaction, and its impact on life in the UV-optical, far-infrared studies that will reveal how habitable planets form, and the examination of the deep cosmos at resolutions that far-exceed those available with single-aperture telescopes.

Our investigation into the feasibility of a lunar-based interferometer represents a significant leap toward establishing larger arrays on the Moon and free-flying in space, both of which circumvent the challenges posed by Earth's atmospheric turbulence. By harnessing a diverse range of wavelengths and delving into various scientific domains, this endeavor promises to unlock new frontiers in lunar and space-based research. The major Vision Mission and Probe studies of the early 2000's, designed for the sub-millimeter out to the X-ray region, ignored the possibility of placing a facility on the Moon if a human infrastructure is present on the Moon. Our study will assess the feasibility of a lunar surface interferometer, which will could provide the basis for future deep space formation-flying interferometers.

This study will teach us much about the drivers, challenges, and solutions to issues involved in the design and construction of interferometers both on the Moon and in space. This information is essential for building the next generation of space interferometers, whether as free-flyers or on the Moon. Overall, this effort will remind people that we can do great things, especially if we work hard together. Our project and bold vision will inspire and excite generations of future Science Technology Engineering Art Mathematics (STEAM) workers.

## ACKNOWLEDGEMENTS

This work is supported in part by an award for proposal 23-NIAC24-B-0060, "A Lunar Long-Baseline Optical Imaging Interferometer: Artemis-enabled Stellar Imager (AeSI)" from the NASA Innovative Advanced Concepts (NIAC) program.